\documentclass[final,1p,times,number]{elsarticle}

\usepackage{amsmath}
\usepackage{amssymb}
\usepackage{amsfonts}
\usepackage{graphicx}
\usepackage{epstopdf}
\usepackage{color}
\usepackage{bm}
\usepackage{multirow}
\usepackage{natbib}
\usepackage{hyperref}

\usepackage{ifpdf}
\ifpdf%
\usepackage{pdflscape}
\else
\usepackage{lscape}
\fi

\DeclareMathOperator{\Tr}{Tr}

\DeclareMathOperator{\sgn}{sgn} 
\DeclareMathOperator{\Imag}{Im}
\DeclareMathOperator{\im}{Im}
 
\DeclareMathOperator{\re}{Re}

\journal{Annals of Physics}
\biboptions{sort&compress}

\begin{document}

\begin{frontmatter}
\title{Entanglement entropy and particle number cumulants of disordered fermions}
\author[ISB1,ISB2]{I.S. Burmistrov\corref{cor}}
\cortext[cor]{Corresponding author. Fax: +7-495-7029317 } 
\ead{burmi@itp.ac.ru}
\author[ISB1,ISB2]{K.S. Tikhonov}
\author[IVG1,IVG2,IVG3]{I.V. Gornyi}
\author[IVG1,IVG2,ADM1]{A.D. Mirlin}
\address[ISB1]{L.D. Landau Institute for Theoretical Physics,
Kosygina street 2, 119334 Moscow, Russia}
\address[ISB2]{\mbox{Laboratory for Condensed Matter Physics, National Research University Higher School of Economics, 101000 Moscow, Russia}}
\address[IVG1]{Institut f\"ur Nanotechnologie,  Karlsruhe Institute of Technology,
76021 Karlsruhe, Germany}
\address[IVG2]{\mbox{Institut f\"ur Theorie der kondensierten Materie,  Karlsruhe Institute of
Technology, 76128 Karlsruhe, Germany}}
\address[IVG3]{A. F.~Ioffe Physico-Technical Institute,
194021 St.~Petersburg, Russia}
\address[ADM1]{Petersburg Nuclear Physics Institute, 188300, St.Petersburg, Russia}

\begin{abstract}
We study the entanglement entropy and particle number cumulants for a system of disordered noninteracting fermions in $d$ dimensions. We show, both analytically and numerically,  that for a weak disorder the entanglement entropy and the second cumulant (particle number variance) are proportional to each other with a universal coefficient. The corresponding expressions are analogous to those in the clean case but with a logarithmic factor regularized by the mean free path rather than by the system size. We also determine the scaling of higher cumulants by analytical (weak disorder) and numerical means. Finally, we predict that  the particle number variance and the entanglement entropy are nonanalytic functions of disorder at the Anderson transition. 
\end{abstract}


\end{frontmatter}

\section{Introduction}

The entanglement entropy of  many-body quantum systems has been attracting a great deal of interest during the last decade. A particularly extensive body of theoretical and numerical studies of entanglement was devoted to translationally invariant interacting quantum systems in $d=1$ spatial dimension, see Refs. \cite{Latorre2009,Eisert2010,Laflorencie2016} for a review. It was found that the entanglement entropy is a useful tool to characterize different phases and to detect quantum phase transitions between them.  The progress  in understanding the entanglement of interacting quantum many-body systems in higher dimensions, $d>1$, was more modest. 

A paradigmatic model for study of the entanglement entropy is a system of noninteracting fermions. It is known \cite{Klich2006,Li2006,Swingle2010} that in the absence of disorder and interaction the entanglement entropy of fermions which fill a Fermi sea in a spatial volume $\sim L^d$ is proportional  to $(k_F L)^{d-1} \ln (k_F L)$ for the case $k_F L \gg 1$, where $k_F$ is the Fermi momentum. A numerical coefficient depends on a geometric shape of the region in the real space as well as on a form of the Fermi surface. Interestingly, the variance of the number of particles in $d$ dimensions depends on the parameter $k_F L$ in a similar way  \cite{Swingle2012}. In fact, this resemblance between the number of particles variance and the entanglement entropy  is not occasional. For noninteracting fermions the latter can be expressed via the very same eigenvalues of wave functions overlaps as the full counting statistics of the number of particles \cite{Klich2006,Klich2009,Peschel2009,Turner2010}. This allows to express the entanglement entropy as a sum over the particle-number cumulants of even orders \cite{Klich2009}. 

The logarithmic enhancement of the entanglement entropy and of the particle-number variance for noninteracting fermions  in comparison with the area law is related with a sharp Fermi surface and with ballistic propagation of particles on spatial scales between the Fermi length $\lambda_F$ and the system size $L$. In the presence of disorder with a mean free path $l\ll L$,  the ballistic motion is possible only at length scales between $\lambda_F$ and $l$. Thus, one may expect that the infrared logarithmic divergence in the expressions for the entanglement entropy and the particle-number variance will be regularized by the mean free path rather than $L$. Therefore, it is natural that the entanglement entropy and the number of particles variance for disordered noninteracting fermions are proportional to $(k_F L)^{d-1}$, i.e. obey the area law. Indeed, the area law for the entanglement entropy was shown recently in the localized phase \cite{Pastur2014}, with an upper boundary for the prefactor related to the localization length. In Ref.  \cite{Potter2014} a behavior of the type $(k_F L)^{d-1} \ln (k_F l)$ (with undefined numerical coefficients) was found for the entanglement entropy and for the particle-number variance. The area-law scaling  of the entanglement entropy was confirmed numerically in Ref. \cite{Pouranvari2015}.

In the presence of disorder, fermions can undergo an Anderson transition from a metallic to an insulating phase. Some aspects of the behavior of the entanglement entropy near the Anderson transition have been addressed in recent works.
In particular, for the problem of a {\it single} fermion in a random potential, it was found that the entropy at the Anderson transition is determined by properties of the singularity spectrum which characterizes the scaling of moments of the wave functions \cite{Kopp2007,Gruzberg2008}. In $d=1$ dimension the Anderson transition can occur in the presence of interaction. In this case the logarithmic dependence of the entanglement entropy on the system size was found to saturate at a scale of the order of the localization length \cite{Berkovits2012,Zhao2013,Berkovits2015}.

In this paper we study the entanglement entropy and moments of the number of particles in a system of disordered noninteracting fermions which fill the Fermi sea at zero temperature. We combine analytical (for weak disorder and near the Anderson transition) and numerical approaches. Our key results are as follows:
\begin{itemize}
\item[(i)]  In the weak-disorder regime, $\lambda_F \ll l \ll L$, the particle-number variance and the entanglement entropy 
are proportional to $(k_F L)^{d-1} \ln (k_F l)$. We determine the numerical factors in these expressions and establish a universal relation between these two quantities, see Eqs. \eqref{eq:PNV:ball} and \eqref{eq:SvsC2}. These analytical results are supported by numerical simulations.
\item[(ii)] We determine, both analytically and numerically, the scaling of higher cumulants of the number of particles.
\item[(iii)] Near the Anderson transition, the particle-number variance and the entanglement entropy obey the area law with a factor which depends in a nonanalytic way on the distance to the critical point, see Eqs. \eqref{eq:NPV:3} and \eqref{eq:EE:3}.
\end{itemize}

The paper is organized as follows. In Sec. \ref{Sec.Form} we remind the general formalism for calculation of the particle-number variance in noninteracting electron system. The results for the  variance in a dirty metal are derived in Sec. \ref{Sec.NPV.DM}. In Sec. \ref{Sec.NPV.A}, the scaling of the particle-number variance near the Anderson transition is studied. In Sec. \ref{Sec.Ent} we explore the behavior of the particle-number cumulants and their relation to the entanglement entropy. Numerical results for the cumulants of the number of particles and for the entanglement entropy are presented in Sec. \ref{Sec.Num}. We end the paper with conclusions, Sec. \ref{Sec.Conc}. Technical details are given in Appendices.

\section{Formalism\label{Sec.Form}}

We start from the standard Hamiltonian describing noninteracting fermions in $d$ dimensions in the presence of a random potential $U(\bm{r})$:
\begin{equation}
\mathcal{H} = \int d^d \bm{r}\, \psi^\dag(\bm{r}) \left [-\frac{\nabla^2}{2m} + U(\bm{r}) \right ] \psi(\bm{r}) .
\label{eq:Ham}
\end{equation}
Here $m$ denotes a particle mass and $\psi^\dag$, $\psi$ stand for the creation and annihilation operators. For the sake of simplicity, we will ignore the electron spin in calculations below. The operator of the number of particles in a given volume $V_L$ is defined as follows
\begin{equation}
\hat{N}_L = \int \limits_{V_L} d^d \bm{r}\,  \psi^\dag(\bm{r})\psi(\bm{r}) ,
\end{equation}
where the spatial integration is restricted to the domain $\bm{r}\in V_L$. We assume that the volume $V_L$ is not isolated but rather is a part of a system of much larger size. Then the number of particles in the volume $V_L$ is a fluctuating quantity, and its variance averaged over disorder realizations can be written as
\begin{gather}
\overline{\langle \langle \hat N^2_L \rangle \rangle} = 
\int \limits_{V_L} d^d \bm{r}
\int \limits_{V_L} d^d \bm{r^\prime} \int \limits_{-\infty}^\infty d\omega\,  dE \, dE^\prime\,
 n_F(E^\prime) \bigl [ 1- n_F(E) \bigr ] \delta(E^\prime-E-\omega) F(E,E^\prime; \bm{r},\bm{r^\prime}) .
 \label{eq:NPV:1}
 \end{gather}
Here the Fermi distribution, $n_F(E) = 1/[1+\exp((E-\mu)/T)]$, is parametrized by the temperature $T$ and the chemical potential $\mu$, and the bar denotes the disorder averaging. The dynamical structure factor $F(E,E^\prime; \bm{r},\bm{r^\prime})$ can be conveniently expressed in terms of the exact eigenfunctions, $\phi_\alpha(\bm{r})$, and eigenenergies, $\varepsilon_\alpha$, of the Hamiltonian \eqref{eq:Ham}:
 \begin{gather}
 F(E,E^\prime; \bm{r},\bm{r^\prime}) = 
 \sum_{\alpha\beta}\overline{\phi^*_\alpha(\bm{r})\phi_\beta(\bm{r})\phi_\alpha(\bm{r^\prime})
 \phi^*_\beta(\bm{r^\prime}) \delta(E-\varepsilon_\beta)\delta(E^\prime-\varepsilon_\alpha)}  .
 \label{eq:F:1}
 \end{gather}

In what follows, we concentrate on the case of the zero temperature, $T=0$. Then, Eq. \eqref{eq:NPV:1} reduces to the following expression:
\begin{gather}
\overline{\langle \langle \hat N^2_L \rangle \rangle} = \int\limits_0^\infty d \omega\, \omega \int \frac{d^d\bm{q}}{(2\pi)^d}\, \mathcal{J}^2_L(q)\,  F(E_F,E_F+\omega,q) ,
\label{eq:PNV:2}
\end{gather}
where $E_F$ stands for the Fermi energy. For a sake of simplicity, we assume that the volume $V_L$ is bounded by a sphere of the radius $L$, i.e.
\begin{equation}
\mathcal{J}_L(q) = \int\limits_{V_L} d^d \bm{r}\,  e^{i \bm{q}\bm{r}} \equiv  \int  d^d \bm{r}\, \theta \bigl (L - r)\, e^{i \bm{q}\bm{r}}  ,
 \end{equation}
 where $\theta(x)$ denotes the Heaviside step function. In this case
\begin{equation}
\mathcal{J}_L(q) = \frac{(2\pi)^{d/2} L^d}{(qL)^{d/2}} J_{d/2}(qL) ,
\label{eq:JL}
\end{equation}
where $J_\nu(x)$ denotes the Bessel function.

Before discussing behaviour of the variance of the number of particles in a disordered metal, we remind the reader its behaviour in the absence of disorder. As is well known, for $k_F L \gg 1$ the particle-number variance becomes  \cite{LS-book,Gioev2006,Klich2006}:
 \begin{equation}
 {\langle \langle \hat N^2_L \rangle \rangle}_{\rm cl} = c_d \frac{(k_F L)^{d-1}}{\pi^2} \ln k_F L  .
 \label{eq:PNV:3}
 \end{equation}
The numerical constant $c_d$ depends on the geometry of the spatial region  $V_L$. 
For the considered case of a spherical volume $V_L$ one finds   (see \ref{Sec.NPV.CM})
\begin{equation}
c_d = \frac{S_d S_{d-1}}{(d-1)2^d \pi^{d-1}}. 
\label{eq:cd:CC}
\end{equation}
Here $S_d= 2\pi^{d/2}/\Gamma(d/2)$ stands for the surface of the $d$ dimensional unite sphere.  We note that the logarithm in Eq. \eqref{eq:PNV:3} appears due to integration over momentum $q$ between scales $1/L$ and $k_F$.

\section{Particle-number  variance in a disordered metal with $k_F l\gg 1$\label{Sec.NPV.DM}}

Now let us consider the variance of the number of particles in the presence of a weak disorder. In what follows we assume that the condition $L\gg l  \gg \lambda_F$ is satisfied. We consider the cases of two and three dimensions. For three dimensions the condition $k_F l \gg 1$ implies that the system is on the metallic side far away from the Anderson transition. In the case of two dimensions we assume that the size $L$ is much shorter than the localization length, $L \ll \xi_{\rm loc}= l \exp (\pi k_F l/2)$, i.e., electron states in the volume $V_L$ do not suffer from strong localization. 	

\subsection{Ballistic to diffusion crossover in $d=2$}

We remind that the dynamical structure factor \eqref{eq:F:1} can be related to the imaginary part of the polarization operator (see e.g., Ref. \cite{Lee1985}):
  \begin{gather}
 F(E,E+\omega; \bm{r},\bm{r^\prime})  = \frac{1}{\pi\omega} 
  \im \Pi^R(\omega, \bm{r}-\bm{r^\prime}) .
  \label{eq:fp}
 \end{gather}
We note that this expression is valid provided one neglects energy dependence of the density of states. Thus the disorder-averaged dynamical structure factor for a two-dimensional disordered metal can be read off from expression for the polarization operator (see for example Ref. \cite{Aleiner2001}): 
\begin{equation}
\Pi^R(\omega,q) = \nu_2 \left [ 1 +\frac{i\omega}{\sqrt{q^2v_F^2+(1/\tau-i\omega)^2}-1/\tau}\right ] .
\label{eq:PiR:d2}
\end{equation}
Here $v_F$ and $\tau$ denote the Fermi velocity and the mean free time, respectively. We note that this expression is valid for $|\omega|\ll E_F$ and $q\ll k_F$. In order to regularize the ultraviolet divergence and to be able to use the asymptotic expression \eqref{eq:PiR:d2}, it is convenient to consider the difference $\Delta \overline{\langle \langle \hat N^2_L \rangle \rangle} = \overline{\langle \langle \hat N^2_L \rangle \rangle} - {\langle \langle \hat N^2_L \rangle \rangle}_{\rm cl}$
 in the variance of the number of particles between disordered, $\overline{\langle \langle \hat N^2_L \rangle \rangle}$,  and clean, ${\langle \langle \hat N^2_L \rangle \rangle}_{\rm cl}$, 
 cases:
\begin{gather}
\Delta \overline{\langle \langle \hat N^2_L \rangle \rangle} =
\frac{\nu_2}{\pi} \int\limits_0^\infty d\omega\, \omega \int\frac{d^2\bm{q}}{(2\pi)^2}\,  \mathcal{J}^2_2(qL)
\re \Biggl [ \frac{1}{\sqrt{q^2v_F^2+(1/\tau-i\omega)^2}-1/\tau} 
-  \frac{1}{\sqrt{q^2v_F^2- \omega^2}}\Biggr ] .
\label{eq:N2}
\end{gather}
Here $\nu_2 = m/(2\pi)$ stands for the density of states.
Performing integration over frequency $\omega$, we find
\begin{gather}
\Delta \overline{\langle \langle \hat N^2_L \rangle \rangle} 
\approx \frac{k_F L^2}{\pi l} \int \limits_{0}^{\infty} \frac{dx}{x} J_1^2\left (\frac{x L}{l}\right ) \Bigl [ {f}_2(x) - x \Bigr ] ,
\end{gather}
where the function 
\begin{equation}
f_2(x)=  \re \int\limits_0^{\infty} dy \, \frac{y}{\sqrt{x^2+(1-i y)^2+1}-1} 
\end{equation}
has the following asymptotic behavior
\begin{equation}
f_2(x) =\begin{cases}
x^2 \ln (\sqrt{2}/x),\qquad  & x\ll 1 ,\\
x, \qquad & x \gg 1 .
\end{cases}
\end{equation}
Then, performing integration over $x$, we find
\begin{gather}
\Delta \overline{\langle \langle \hat N^2_L \rangle \rangle} 
= \frac{k_F L}{\pi^2} \Bigl [ -\ln (L/l) + a_2 \Bigr ] ,
\label{eq:bd:2}
\end{gather}
where $a_2$ is a numerical constant  given by
\begin{equation}
a_2  = \int \limits_{0}^{1} \frac{dx}{x^2} f_2(x) +\int \limits_{1}^{\infty} \frac{dx}{x^2} \Bigl [f_2(x)-x \Bigr ] 
- \pi \int \limits_0^1 dx  J_1^2(x) - \int \limits_1^\infty \frac{dx}{x} \Bigl [\pi x  J_1^2(x)  -1 \Bigr ]
\approx  0.8 .
\end{equation}

\subsection{Ballistic to diffusion crossover in $d=3$}

The disorder averaged polarization operator in $d=3$ at small frequencies,
$|\omega|\ll E_F$, and momentum, $q\ll k_F$, can be written as
\begin{equation}
\Pi^R(\omega,q) = \nu_3 \left [ 1 +i\omega \left (\frac{q v_F}{\arctan[q l/(1-i\omega \tau]} -\frac{1}{\tau}\right )^{-1}\right ] .
\end{equation}
Here $\nu_3 = m k_F/(2\pi^2)$ denotes the density of states at the Fermi energy.
The difference in the variance of the number of particles between disordered and clean cases is given as follows
\begin{gather}
\Delta \overline{\langle \langle \hat N^2_L \rangle \rangle} =
\frac{\nu_3}{\pi} \int\limits_0^\infty d\omega\, \omega \int\frac{d^3\bm{q}}{(2\pi)^3}\,  \mathcal{J}^2_3(qL)
 \Biggl [ \re \left (\frac{q v_F}{\arctan[q l/(1-i\omega \tau]} -\frac{1}{\tau}\right )^{-1} - 
  \frac{\pi \theta(q v_F - \omega)}{2q v_F}\Biggr ] .
\label{eq:N3}
\end{gather}
Integrating over $\omega$, we find:
\begin{gather}
\Delta \overline{\langle \langle \hat N^2_L \rangle \rangle} =
\frac{k_F^2 L^3}{2 \pi l} \int \limits_0^\infty \frac{dx}{x} J_{3/2}^2 \left (x \frac{L}{l} \right ) 
\left [ f_3(x) - x \right ] ,
\end{gather}
where the function
\begin{equation}
f_3(x) = \frac{4}{\pi} \re \int\limits_0^{\infty} dy \, y  \left ( \frac{x}{\arctan \frac{x}{1-i y}}-1\right )^{-1}
\end{equation}
has the following asymptotic behavior
\begin{equation}
f_3(x) = \frac{1}{3\pi} \begin{cases}
4x^2 \ln (3/x^2), & x\ll 1 ,\\
3\pi x , & x\gg 1 .
\end{cases}
\end{equation}
Now performing integration over $x$, we obtain
\begin{gather}
\Delta \overline{\langle \langle \hat N^2_L \rangle \rangle} 
= \frac{k_F^2 L^2}{2\pi^2} \Bigl [ -\ln (L/l) + a_3 \Bigr ] ,
\label{eq:bd:3}
\end{gather}
where
\begin{equation}
a_3  = \int \limits_{0}^{1} \frac{dx}{x^2} f_3(x) +\int \limits_{1}^{\infty} \frac{dx}{x^2} \Bigl [f_3(x)-x \Bigr ] 
- \pi \int \limits_0^1 dx  J_{3/2}^2(x) - \int \limits_1^\infty \frac{dx}{x} \Bigl [\pi x  J_{3/2}^2(x)  -1 \Bigr ]
\approx  1.1 .
\end{equation}

\subsection{Diffusive contribution}

The main contribution (proportional to $\ln k_F l$) to the particle-number  variance in a disordered metal with $k_F l\gg 1$ comes from the ballistic scales, i.e. from integration over momentum $q$  in the range between $l^{-1}$ and $k_F$. This can be seen by computing the contribution to the variance from the diffusive region. We use Eq. \eqref{eq:fp} and the expression for the polarization operator in the diffusive approximation: 
\begin{equation}
\Pi^R(\omega,q)= \nu_d \frac{D(q,\omega) q^2}{D(q,\omega) q^2 - i \omega}  ,
\label{eq:PO:2}
\end{equation}
where $\nu_d = S_d k^{d-1}_F/((2\pi)^d v_F)$ denotes the density of states at the Fermi energy. Then, we rewrite the diffusive contribution to the variance of the number of particles as
\begin{gather}
\overline{\langle \langle \hat N^2_L \rangle \rangle}_{\rm diff}  = \frac{\nu_d}{\pi}  \int \limits_{q<\Lambda_q} \frac{d^d \bm{q}}{(2\pi)^d} \mathcal{J}_L^2(q)  \int\limits_0^{\Lambda_\omega} d\omega  \im \frac{D(q,\omega) q^2}{D(q,\omega) q^2 - i \omega} 
\label{eq:PNV:diff1}
\end{gather}
where we introduce two ultraviolet cutoffs: $\Lambda_\omega \sim 1/\tau$ and $\Lambda_q \sim 1/l$. In the metallic case the diffusive coefficient is the constant, $D(q,\omega) \equiv D = v_F^2 \tau/d$. Hence, the diffusive contribution is given as
\begin{gather}
\overline{\langle \langle \hat N^2_L \rangle \rangle}_{\rm diff} \approx  \frac{\nu_d S_d L^{d-1}}{\pi^2}  \int \limits_{L^{-1}}^{\Lambda_q} \frac{dq}{q^2}  \int\limits_0^{\Lambda_\omega} d \omega\, \frac{\omega Dq^2}{(D q^2)^2+\omega^2}  = \beta_d^{\rm diff} \frac{ (k_F L)^{d-1}}{\pi^2}.
\label{eq:PNV:diff1}
\end{gather}
We thus see that the diffusive contribution statisfies the area law.
The numerical constant $\beta_d^{\rm diff}$ depends on the ratio $v_F \Lambda_q/\Lambda_\omega$ since the integrals over 
$q$ and $\omega$ are dominated by the ultraviolet. Thus, $\beta_d^{\rm diff}$ cannot be determined accurately within such calculation. However, we stress that it is independent of the disorder for $k_F l \gg 1$. We note that the result \eqref{eq:PNV:diff1} in the case of $d=3$ has been derived recently in Ref. \cite{Potter2014}.

\subsection{The number of particles variance in a disordered metal}

The results obtained above imply that the variance of the number of particles averaged over disorder realizations  can be written in a metallic case, $L\gg l \gg \lambda_F$, as 
\begin{equation}
\overline{\langle \langle \hat N^2_L \rangle \rangle} = c_d \frac{(k_F L)^{d-1}}{\pi^2} \Bigl (\ln k_F l + {\rm const}\Bigr ).
\label{eq:PNV:ball}
\end{equation}
The numerical coefficient $c_d$ in Eq. \eqref{eq:PNV:ball} is exactly the same as in the clean case, Eq. \eqref{eq:cd:CC}. We note that the result \eqref{eq:PNV:ball} was obtained  for $d=3$  in Ref. \cite{Potter2014}. However, the numerical factor $c_d$ in this formula was not deteremined there.

The result \eqref{eq:PNV:ball} has a transparent physical explanation. As follows from Eqs. \eqref{eq:bd:2}, \eqref{eq:bd:3} and \eqref{eq:PNV:diff1}, the main contribution to the particle-number variance comes from ballistic momentum scales, $1/l \ll q \ll k_F$. For such momenta the variance of the number of particles can be evaluated in the same way as in the absence of disorder. The only difference is in the infrared cutoff for the logarithmic integral over momentum: $1/l$ instead of $1/L$.

\section{Scaling of the particle-number variance near the Anderson transition\label{Sec.NPV.A}}

In this section we consider disordered noninteracting fermions near an Anderson transition. Let us denote by $g_*$ the value of the dimensionless conductance at which the Anderson transition occurs. We are interested in the behaviour of the variance of the number of particles in the critical region, $|g_0 - g_*|/g_* \ll 1$. Here $g_0$ is the bare conductance at the length scale of the order of the mean free path which can be estimated as $g_0 \sim k_F l$. In a 3D system, an Anderson transition occurs at the value of the dimensionless conductance of the order unity, $g_* \sim 1$. Therefore, since we are interested in the critical region where the bare conductance $g_0 \sim g_* \sim 1$, the ballistic contribution to the variance of the number of particles cannot be computed 
accurately. Roughly, it can be estimated by Eq. \eqref{eq:PNV:ball} with $k_F l$ substituted by $g_0$. 
We expect that the ballistic contribution  is a regular function of $g_0$ near the Anderson transition. 
Below we consider the diffusive contribution to the particle-number variance.

\subsection{Exactly at criticality}

At criticality, the diffusion coefficient acquires a frequency and momentum dependence. Exactly, at the critical point, $g_0=g_*$, this dependence can be written in the scaling form as \cite{Wegner1976}: 
\begin{equation}
D(q,\omega) = \frac{g_*}{\nu_d} q^{d-2} \mathcal{F}\left (\frac{\nu_d \omega}{g_* q^d} \right )  .
\end{equation}
Here a regular function $\mathcal{F}(X)$ has the following asymptotic behaviour
\begin{gather}
\mathcal{F}(X) \propto \begin{cases}
X^{-\Delta_2/d} , \quad X\ll  1 , \\
X^{(d-2)/d}, \quad X \gg 1 ,
\end{cases}
\end{gather}
with $\Delta_2$ standing for the multifractal exponent controlling the scaling of the fourth moment of a wave function. 
Using Eq. \eqref{eq:PNV:diff1}, the diffusive contribution to the variance of the  number of particles can be written as
\begin{gather}
\overline{\langle \langle \hat N^2_L \rangle \rangle}_{\rm diff} = \frac{\nu_d S_d L^{d-1} \Lambda_\omega}{ d\pi^2 \Lambda_q} r^{-1/d}
\int \limits_{r}^{\infty} dz\, z^{1/d-2} \int\limits_0^{z} d X\, \frac{X \mathcal{F}(X)}{X^2 + \mathcal{F}^2(X)}
= \alpha_d^{\rm diff} \frac{(k_F L)^{d-1}}{\pi^2} ,
\end{gather}
where  $r = \nu_d \Lambda_\omega/(g_* \Lambda_q^d) \sim 1$. The dimensionless constant 
\begin{equation}
\alpha_d^{\rm diff} = \frac{S^2_d }{(d-1) (2\pi)^{d}}\frac{\Lambda_\omega}{v_F \Lambda_q } r^{-1/d}
 \int \limits_0^{\infty}  d X\, \bigl (\max\{r, X\}\bigr )^{1/d-1} \, \frac{X \mathcal{F}(X)}{X^2 + \mathcal{F}^2(X)} 
 \end{equation}
 is expected to be of order unity since $\Lambda_\omega/ (v_F \Lambda_q) \sim 1$. 
 
Interestingly, at the critical point the diffusive contribution to the number of particles variance  is similar to the diffusive contribution in the metallic phase far away from criticality. However, in the metallic phase with $g_0\gg g_*$ the diffusive contribution is much smaller than the ballistic one whereas at the critical point the diffusive and ballistic contributions are of the same order of magnitude.

\subsection{Slightly off criticality: metallic side}

Away from the critical point the correlation/localization length $\xi$ is finite. Slightly off criticality, $|g_0/g_*-1|\ll 1$, the correlation/localization length is determined by the corresponding critical exponent $\nu$, $\xi = l |g_0/g_*-1|^{-\nu}$. 
The frequency and momentum dependence of the diffusion coefficient on the metallic side from the Anderson transition reads \cite{Wegner1976}:
\begin{equation}
D(q,\omega) = (g_*/\nu_d){\xi}^{2-d} \mathcal{R}(\omega/\Delta_\xi,q\xi) ,
\end{equation}
where $\Delta_\xi= g_*/(\nu_d\xi^{d})$ stands for the mean single-particle level spacing in the volume $\xi^d$. A regular function $\mathcal{R}(\Omega,Q)$ has the following asymptotic behaviour
\begin{equation}
\mathcal{R}(\Omega,Q)
\propto \begin{cases}
1, & |\Omega|\ll 1, \, Q\ll 1 , \\
|\Omega|^{(d-2)/d} , & |\Omega| \gg 1, \, Q \ll |\Omega|^{1/d} , \\
Q^{d-2+\Delta_2} |\Omega|^{-\Delta_2/d} , & |\Omega| \gg 1, \, Q \gg |\Omega|^{1/d} , \\
Q^{d-2+\Delta_2} , & |\Omega| \ll 1, \, Q \gg 1 .
\end{cases}
\end{equation}
It is convenient to choose normalization of the function $\mathcal{R}(\Omega,Q)$ in such a way that 
$\mathcal{F}\bigl (\nu_d\omega/(g_* q^d)\bigr ) = \lim \limits_{\xi \to \infty} Q^{2-d} \mathcal{R}(\Omega,Q)$. Then, for $L\gg \xi$ the diffusive contribution to the particle-number variance becomes 
\begin{gather}
\overline{\langle \langle \hat N^2_L \rangle \rangle}_{\rm diff}  = \alpha^{\rm diff}_d \frac{(k_F L)^{d-1}}{\pi^2} + {\beta^{\rm met}_d} \left ( \frac{L}{\xi}\right )^{d-1},
\label{eq:PNV:ms1}
\end{gather}
where 
\begin{gather} 
\beta^{\rm met}_d= \frac{g_* S_d}{\pi^2} \int \limits_0^{\infty}  \frac{dQ}{Q^2} \int\limits_0^{\infty} d\Omega \Biggl \{ \frac{\Omega Q^2 \mathcal{R}(\Omega,Q)}{[Q^2 \mathcal{R}(\Omega,Q)]^2+\Omega^2} - \frac{\Omega Q^d \mathcal{F}(\Omega/Q^d)}{[Q^d \mathcal{F}(\Omega/Q^d)]^2+\Omega^2} \Biggr \} .
\end{gather}
The numerical coefficient $\beta^{\rm met}_d$ is expected to be of the order unity and is determined by the behaviour of the diffusion coefficient  on the metallic side of the critical region of the Anderson transition. 

\subsection{Slightly off criticality: insulating side}

The frequency and momentum dependence of the diffusion coefficient on the insulating side from the Anderson transition should be consistent with the Mott's formula \cite{Mott1968,Mott1970}. We thus find \cite{Burmistrov2014}:
\begin{equation}
\mathcal{R}(\Omega,Q)
\propto \begin{cases}
-i\Omega+c \Omega^2 \ln^{d+1}(1/|\Omega|), & |\Omega|\ll 1, \, Q\ll 1 ,\\
|\Omega|^{(d-2)/d} , & |\Omega| \gg 1, \, Q \ll |\Omega|^{1/d} ,\\
Q^{d-2+\Delta_2} |\Omega|^{-\Delta_2/d} , & |\Omega| \gg 1, \, Q \gg |\Omega|^{1/d} ,\\
Q^{d-2+\Delta_2}(-i\Omega+c \Omega^2 \ln^{d+1}(1/|\Omega|)) , & |\Omega| \ll 1, \, Q \gg 1 .
\end{cases}
\end{equation}
Using this formula, we find the diffusive contribution to the variance of the number of particles for a system size exceeding the localization length, $L\gg \xi$,
\begin{gather}
\overline{\langle \langle \hat N^2_L \rangle \rangle}_{\rm diff}  = \alpha^{\rm diff}_d \frac{(k_F L)^{d-1}}{\pi^2} + {\beta^{\rm ins}_d} \left ( \frac{L}{\xi}\right )^{d-1},
\label{eq:PNV:is1}
\end{gather}
where 
\begin{gather} 
\beta^{\rm ins}_d= \frac{g_* S_d}{\pi^2} \int \limits_0^{\infty}  \frac{dQ}{Q^2} \int\limits_0^{\infty} d\Omega \Biggl \{ \frac{\Omega Q^2 \re \mathcal{R}(\Omega,Q)}{[Q^2 \re \mathcal{R}(\Omega,Q)]^2+[\Omega-Q^2 \im \mathcal{R}(\Omega,Q)]^2}- \frac{\Omega Q^d \mathcal{F}(\Omega/Q^d)}{[Q^d \mathcal{F}(\Omega/Q^d)]^2+\Omega^2} \Biggr \} .
\end{gather}
The numerical coefficient $\beta_d^{\rm ins}$ is expected to be of the order unity and is determined by the behaviour of the diffusion coefficient in the critical region on the insulating side. 

We note that the results \eqref{eq:PNV:ms1} and \eqref{eq:PNV:is1} can be guessed on the basis of the following suppositions. First, the particle-number variance in the critical region scales with the system size as $L^{d-1}$ (area law) and is finite at the critical point. Second, in addition to $L$ there is a single length scale---the correlation length $\xi$---that determines the behaviour of  deviation of the area-law prefactor from its value at the critical point.  This leads to Eqs.~\eqref{eq:PNV:ms1} and \eqref{eq:PNV:is1}.

The results \eqref{eq:PNV:ms1} and \eqref{eq:PNV:is1} imply that in the critical region of Anderson transition, $|g_0/g_*-1|\ll 1$, the particle-number variance can be written as
\begin{equation}
\overline{\langle \langle \hat N^2_L \rangle \rangle} =
 \left ( \frac{L}{l}\right )^{d-1} 
 \begin{cases}
 \alpha_d  + \tilde{\beta}^{\rm met}_d (g_0-g_*)^{\nu(d-1)}, & \qquad g_0 \geqslant g_* , \\
\alpha_d + \tilde{\beta}^{\rm ins}_d (g_*-g_0)^{\nu(d-1)}, & \qquad g_0 < g_*  ,
\end{cases}
\label{eq:NPV:3}
\end{equation}
where $\alpha_d$ is a regular function of $g_0-g_*$. The numerical coefficients $\tilde{\beta}^{\rm met}_d$ and $\tilde{\beta}^{\rm ins}_d$ are related in an obvious way to the numerical coefficients ${\beta}^{\rm met}_d$ and ${\beta}^{\rm ins}_d$. 

The result \eqref{eq:NPV:3} implies a nonanalytic behaviour of $\overline{\langle \langle \hat N^2_L \rangle \rangle}$ as a function of disorder, $g_0$, in the vicinity of the critical point $g_*$. For Anderson transitions in two and three dimensions the exponent $\nu (d-1)$ is larger than $2$. Thus a singular behaviour of the particle-number variance manifests itself only  in its sufficiently high derivatives with respect to $g_0$.

\section{Scaling of the entanglement entropy and of particle-number cumulants \label{Sec.Ent}}

We remind the reader  that the entanglement entropy $S_L$ of a volume $V_L$ is defined via the reduced density matrix $\rho_L = \Tr_{\underline{V_L}} \rho$, where the trace is taken over the states in the region $\underline{V_L}$ which is complementary to $V_L$:
\begin{equation}
S_L = - \Tr \rho_L \ln \rho_L  .
\end{equation} 
In the noninteracting case the entanglement entropy $S_L$ can be expressed as follows \cite{Klich2006,Klich2009,Peschel2009,Turner2010}
\begin{equation}
S_L= - \sum\limits_j \Bigl [\lambda_j \ln \lambda_j + (1-\lambda_j) \ln (1-\lambda_j) \Bigr ] ,
\label{eq:EE:L1}
\end{equation}
where  $0< \lambda_j < 1 $ denotes eigenvalues of the following single-particle correlation function
\begin{equation}
\Phi_L(\bm{r}, \bm{r^\prime}) = \sum\limits_\alpha \phi^*_\alpha(\bm{r}) \phi_\alpha(\bm{r^\prime})\delta(E_F-\varepsilon_\alpha), \qquad \bm{r}, \bm{r^\prime} \in V_L .
\end{equation}
The variance of the number of particles at zero temperature can be expressed via the same eigenvalues:
\begin{equation}
{\langle \langle \hat N^2_L \rangle \rangle} = \sum\limits_j \lambda_j(1-\lambda_j) .
\label{eq:EE:L2}
\end{equation} 
The particle-number variance provides the low bound for the entanglement entropy, $S_L \geqslant 4 \ln 2\, \langle \langle \hat N^2_L \rangle \rangle$ \cite{Klich2006}.

The set of eigenvalues $\lambda_j$ determines the characteristic function of the full counting statistics of the number of particles in the volume $V_L$ \cite{Levitov1993}:
 \begin{equation}
 \chi(\theta) = \left \langle e^{i \theta \hat N_L } \right \rangle  = \prod \limits_j \Bigl ( 1 - \lambda_ j + e^{i\theta} \lambda_j \Bigr ) .
 \end{equation}
This characteristic function encodes information about fluctuations of the number of particles and can be expressed via cumulants  $C_{m} = \Bigl \langle \bigl (\hat N_L - \langle \hat N_L \rangle \bigr )^{m} \Bigr \rangle$:
\begin{equation}
\ln\chi(\theta) = \sum\limits_{m=1}^\infty \frac{(i\theta)^m C_m}{m!} .
\end{equation}
In Ref. \cite{Klich2009}, these relations have been used in order to express the entanglement entropy
as the following infinite series:
\begin{equation}
S_L = 2 \sum_{m=1}^\infty \zeta (2m) C_{2m}\,  .
\label{eq:SvsC}
\end{equation}
Here $\zeta(m)$ denotes the Riemann zeta function.

The results obtained recently in Ref. \cite{Ivanov2013}  for the full counting statistics in the absence of disorder for $d=1$ imply that the cumulants $C_{2m}$ with $m\geqslant 2$ are independent of $k_F L$
provided $L\gg \lambda_F$. Therefore, the leading behaviour of the entanglement entropy for $k_F L \gg 1$ in $d=1$ is determined by the particle-number variance only:
\begin{equation}
S_L \simeq \frac{\pi^2}{3} C_2  .
\label{eq:SvsC2}
\end{equation} 

For arbitrary dimension $d$, the cumulants  $C_{2m}$ with $m\geqslant 2$ in the clean case are calculated in \ref{app}. The result reads 
\begin{equation}
C_{2m} = a_{2m}^{(d)} (k_F L)^{d-1}, \qquad  k_F L\gg 1 .
\label{eq:C2m:arb:d}
\end{equation}
Thus, contrary to the variance, higher cumulants ($m\ge 2$) obey a conventional area law, without a logarithmic enhancement. Therefore, the variance ($m=1$) is larger, due to the logarithmic factor, than higher cumulants, so that Eq. \eqref{eq:SvsC2} is valid for an arbitrary dimension $d$ in the absence of disorder \cite{Flindt2011,Calabrese2012,Flindt2012}. We note a subtle point related to summation of the series in Eq. \eqref{eq:SvsC}. Specifically, the coefficients $a_{2m}^{(d)}$ grow factorially, e.g., $|a_{2m}^{(d=1)}| \sim (2m)!/[(2\pi^2)^m m!]$ \cite{Ivanov2013}, so that the series is formally not convergent.  However, since the even cumulants  have alternating signs, the series \eqref{eq:SvsC} can be summed up using the integral representation of the zeta-function (see \ref{AppSL}). Alternatively, one can derive another expansion of the entanglement entropy in terms of the even cumulants,  which yields a series that can be summed up directly \cite{Flindt2011,Calabrese2012,Flindt2012,Flindt2015}. As shown in \ref{AppSL}, both ways lead to equivalent results. This confirms the validity of Eq.~\eqref{eq:SvsC2}, with the total contribution of higher cumulants being smaller by a logarithmic factor.

In the case of a dirty metal, the disorder-averaged cumulants with $m\geqslant 2$ have the same dependence on the size $L\gg l \gg \lambda_F$ as for the clean case  (see \ref{app}):
\begin{equation}
 \overline{C_{2m}} = \tilde{a}_{2m}^{(d)} (k_F L)^{d-1}, \qquad k_F^{-1} \ll l \ll L .
 \label{eq:C2m:dis-aver}
 \end{equation} 
The validity of Eq. \eqref{eq:SvsC2} is supported also by numerical calculations presented in the next section. Also, our numerics demonstrate that coefficients $\tilde{a_{2m}}^{(d)}$ with $m\geqslant 2$ tend to constants at $k_F l \gg 1$, in consistency with the clean-limit behavior.

Therefore, the results of the previous section for the particle-number variance in a dirty metal can be directly applied to the disorder-averaged entanglement entropy $\overline{S_L}$.  In particularly, this implies that the disorder-averaged entanglement entropy obeys the area law, $\overline{S_L} \sim (k_F L)^{d-1}$, in agreement with the results of Refs. \cite{Pastur2014,Potter2014}. Moreover, since the second cumulant $\overline{C_2}$ is parametrically larger than the other ones, $\overline{C_4}, \overline{C_6}, \dots $, for $k_F l \gg 1$, Eq. \eqref{eq:SvsC2} remains valid for the disordered averaged quantities in this case.

Near the critical point of the Anderson transition, the second cumulant $\overline{C_2}$ has a non-analytic behaviour controlled by the critical scaling of the correlation/localization length $\xi$ (see Eqs. \eqref{eq:PNV:ms1} and \eqref{eq:PNV:is1}). As outlined in the end of Sec. \ref{Sec.NPV.A}, 
these results can be anticipated on the basis of simple qualitative arguments. First, the number of particle variance scales with the system size as $L^{d-1}$  in the critical region and is finite at the critical point. Second, the deviation of the corresponding prefactor from its value at the critical point is determined only by the single length scale, $\xi$.  These arguments are expected to apply to the other cumulants $C_{2m}$ as well. Therefore,  
in analogy with \eqref{eq:NPV:3}, the disorder-averaged entanglement entropy should have a nonanalytic dependence on the bare dimensionless conductance at the Anderson criticality, $|g_0/g_*-1|\ll 1$:  
\begin{equation}
\overline{S_L} =
 \left ( \frac{L}{l}\right )^{d-1} 
 \begin{cases}
 \check{\alpha}_d  + \check{\beta}^{\rm met}_d (g_0-g_*)^{\nu(d-1)}, & \qquad g_0 \geqslant g_* , \\
\check{\alpha}_d + \check{\beta}^{\rm ins}_d (g_*-g_0)^{\nu(d-1)}, & \qquad g_0 < g_*  ,
\end{cases}
\label{eq:EE:3}
\end{equation}
where $\check{\alpha}_d$ is a regular function of $(g_0-g_*)$, and $\check{\beta}^{\rm met}_d$ and  $\check{\beta}^{\rm ins}_d$ are some constants. Since in the case of the Anderson transitions in two and three dimensions the exponent $\nu (d-1)$ is larger than $2$, the non-analytic behaviour of $\overline{S_L}$  can be observed in its high derivatives with respect to $g_0$ only.

\section{Particle-number cumulants and entanglement entropy:  Numerical results \label{Sec.Num}}

In this section, we supplement our analytical findings by numerical simulations.
We recall that, in view of Eq. (\ref{eq:SvsC}), the properties of entanglement
entropy $S_{L}$ can be straightforwardly inferred from those of the
cumulants $C_{2m}$. In the preceding Section, we have shown that in a
disordered metal, higher-order ($m\geq 2$) cumulants scale with $L$ in the same way as $C_{2}$, but are smaller (no logarithmic enhancement) in the limit of $k_F l\gg1$. In this section we check this result numerically for 2D Anderson model. In particular, we study the size-dependence of the first few cumulants and confirm Eq. \eqref{eq:C2m:dis-aver}. As we have mentioned above, this implies the remarkable identity (\ref{eq:SvsC2}), which we explicitly check numerically as well.

We study non-interacting spinless particles hopping over a 2D square lattice with periodic boundary conditions in a potential disorder described by the Hamiltonian
\begin{equation}
\label{H}
\mathcal{H}=-\sum_{\left<i, j\right>}\left(a_i^+ a_j + a_j^+ a_i\right)+\sum_{i} \epsilon_i a_i^+ a_i\,,
\end{equation}
where the first sum is over the nearest-neighbour sites of the lattice. The energies $\epsilon_i$ are independent random variables sampled from a uniform distribution on $[-W/2,W/2]$. All states in this model are known to be localized, but in the middle of the band the localization length is exponentially large at moderate $W$. In this situation, effects of strong localization are immaterial for not too large $L$ that can be studied in numerical simulations, and the treatment of disorder carried out above should be sufficient. On the other hand, for strong disorder (large $W$), the strong localization effects should manifest themselves already in relatively small systems. 

\begin{figure}
\centerline{\includegraphics[width=0.9\linewidth]{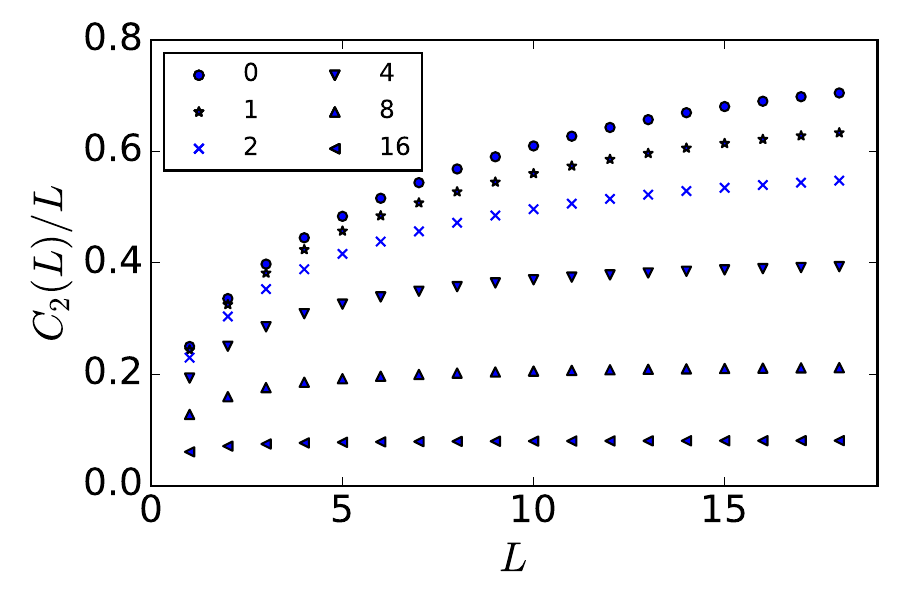}}
\caption{Size-dependence of the cumulant $C_2$ of a 2D disordered system. For a series of disorder values (see legend for $W$), the value of $C_2(L)/L$ is shown. Note the unbounded growth of $C_2(L)/L$ with increase of $L$ at $W \to 0$.}
\label{fig1}
\end{figure}
\begin{figure}
\centerline{\includegraphics[width=0.9\linewidth]{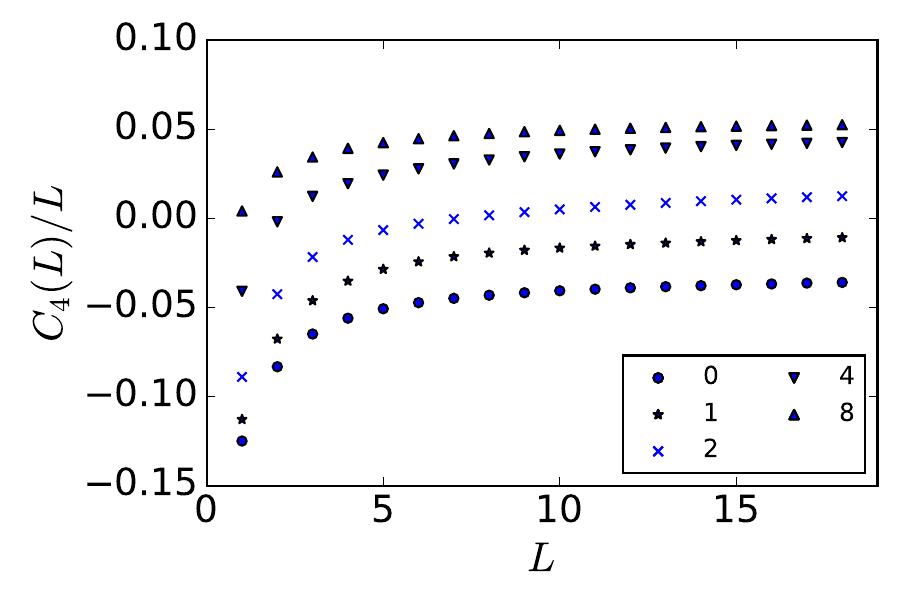}}
\caption{Size-dependence of the cumulant $C_4$. For a series of disorder values (see legend for $W$), the value of $C_4(L)/L$ is shown. Note that $C_4(L)/L$ tends to saturate in the limit of $W\to 0$ and remains bounded at all other disorder values.}
\label{fig2}
\end{figure}

Let us first consider the cumulant $C_2$. The disorder-averaged ratio $C_2/L$ is shown on the Fig. \ref{fig1} for several values of disorder parameter $W$ as a function of the system size. Note that weak (logarithmic) growth of $C_2(L)/L$ for relatively small systems is succeeded by saturation at a finite disorder-dependent value. This is in line with our expectations and corresponds to ballistic-diffusive crossover from Eq. (\ref{eq:PNV:3}) to Eq. (\ref{eq:PNV:ball}).

\begin{figure}
\centerline{\includegraphics[width=0.9\linewidth]{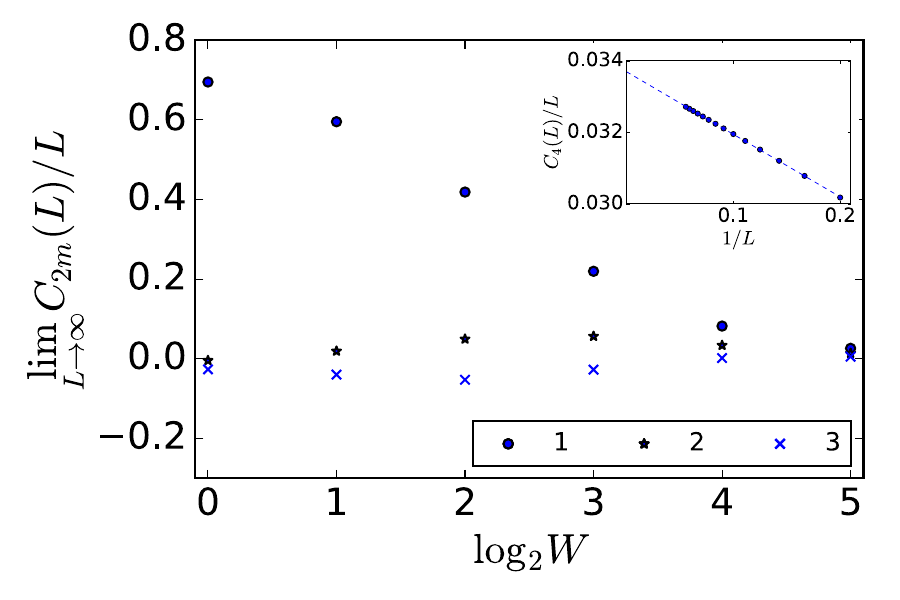}}
\caption{Disorder-dependence of the cumulant $C_{2m}(L)/L$ at $L\to \infty$ for $m=1,2,3$ (see legend for $m$). Note the unbounded growth of $\lim \limits_{L\to \infty} C_2(L)/L$ at $W\to 0$ and finiteness of 
$\lim \limits_{L\to \infty} C_{4,6}(L)/L$. The inset shows $C_4(L)/L$ as a function of $1/L$ for $W=16$.}
\label{fig3}
\end{figure}

A similar analysis can be performed for higher cumulants, with the results for $m=2$ (all other higher cumulants behave in a qualitatively similar way) shown in Fig. \ref{fig2}. In this figure, the values of the ratio $C_{4}(L)/L$ are shown for various system sizes. The saturation at a disorder-dependent value is manifested in these plots. In Fig. \ref{fig3} we present the disorder dependence of the saturation value of the ratio $\lim \limits_{L\to \infty} C_{2m}(L)/L$ for $m=1,2,3$. These numerical results demonstrate clearly our key observation: unbounded growth of $\lim \limits_{L\to \infty} C_{2}(L)/L$ as $W\to 0$ and boundedness of $\lim \limits_{L\to \infty} C_{m>2}(L)/L$ for all disorder values. In the inset to Fig. \ref{fig3}  we illustrate that in the case of strong disorder the ratio $C_{4}(L)/L$ approaches the constant as a linear function of $1/L$.

\begin{figure}
\centerline{\includegraphics[width=\linewidth]{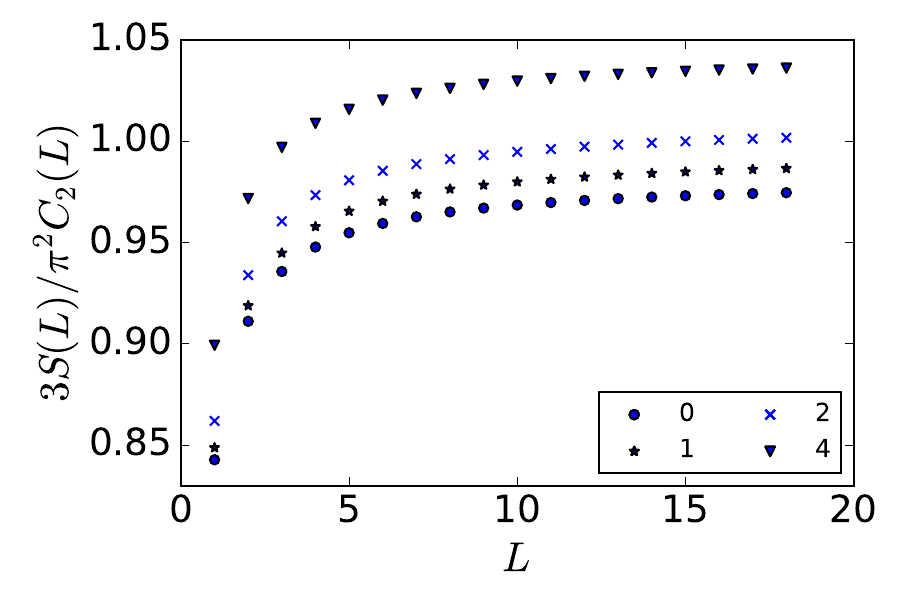}}
\caption{Size-dependence of $3 S_L/[\pi^2 C_2(L)]$ for different values of disorder $W$ (see legend for $W$).}
\label{fig4}
\end{figure}

Finally, we verify numerically the relation, Eq. (\ref{eq:SvsC2}), between the entanglement entropy and the particle-number variance. In order to do so, we plot in Fig. \ref{fig4} the ratio $3 S_L/[\pi^2 C_2]$ at not too high disorder values and for various system sizes. We observe a saturation of this ratio at a value fairly close to the unity, as expected from our analysis for the case of weak disorder. The saturation takes place at system sizes consistent with ballistic-diffusive crossover observed in Fig.  \ref{fig1}. In the clean case, the value of the ratio $3 S_L/[\pi^2 C_2]$ remains slightly below (around $3\%$) for our largest system size, $L=18$. The approach to the unity for the clean system is quite slow, in agreement with its analytically expected logarithmic character.  It is interesting that the limiting value of $3 S_L/[\pi^2 C_2]$  deviates only weakly (a few percent) from the unity even for quite strong disorder ($W=4$), which is a manifestation of a numerical smallness of higher cumulants, see Fig.~\ref{fig3}. 

It is worth emphasizing that, in the weak-disorder regime, the prefactor in the area law for the entropy and the number variance is controlled by the mean free path $l$ and not by the localization length $\xi$ (which is much larger in a 2D system). Therefore, in this situation, the upper boundary for this prefactor found in Ref.~\cite{Pastur2014} is totally different from its actual value. 

\section{Conclusions\label{Sec.Conc}}

To summarize, we have studied the behaviour of the particle-number cumulants and of the entanglement entropy for a $d$-dimensional system of noninteracting fermions  in the presence of disorder at zero temperature. All of these quantities were found to obey the area law. We have shown that for a weak disorder the entanglement entropy and the second cumulant (particle number variance) are proportional to each other with a universal coefficient, see Eq. \eqref{eq:SvsC2}. The corresponding expressions for both quantities are analogous to those in the clean case but with a logarithmic factor regularized by the mean free path rather than by the system size, i.e. $\ln k_F L$ replaced by $\ln k_F l$, see Eq. \eqref{eq:PNV:ball}. Higher cumulants do not show the logarithmic enhancement for weak disorder. We have also shown that the particle-number cumulants and the entanglement entropy have a non-analiticity at the point of the Anderson transition. This non-analiticity is controlled by the exponent $\nu (d-1)$ where $\nu$ is the localization length exponent, see Eqs. \eqref{eq:NPV:3} and \eqref{eq:EE:3}. Our theoretical results are supported by numerical calculations of the cumulants and of the entanglement entropy for 2D disordered systems. 

Before closing the paper, we briefly discuss possible extensions of our result. One of interesting directions is to study the entanglement entropy and the particle-number cumulants in a disordered fermionic system at finite temperature $T$. In particular, it would be interesting to see what is a fate  of the relation between the particle-number variance and the entanglement entropy at finite $T$. 

The second, and much richer, direction for expected extension of our results is inclusion of the electron-electron interaction in addition to disorder. A natural guess is that the weak-disorder behavior of the entropy and of number cumulants (area law, with logarithmic enhancement of the entanglement entropy and of the number variance in the weak-disorder regime) remains qualitatively the same as long as the system is in the Fermi-liquid phase. On the other hand, it is a priori not clear whether the relation between the entropy and the variance should survive. In fact, the arguments presented in Ref.~\cite{Swingle2012} for a clean system suggest that this relation does not survive, since the entropy is affected by the Landau interaction parameter, while the variance is not. A systematic analysis in the presence of disorder remains to be carried out. 

Of particular interest is the behavior near the localization transition in the presence of interaction.  The structure of the polarization operator of an interacting system in the diffusive regime remains   the same as in the noninteracting case. Therefore, we expect that the nonanalytic contribution to $\overline{\langle \langle \hat N^2_L \rangle \rangle}$ at the critical region of the Anderson transition survives in the presence of electron-electron interaction. Interestingly, the behaviour of the entanglement entropy with increase of disorder consistent with Eq. \eqref{eq:EE:3} was recently observed  in numerical simulations of disordered interacting electrons at filling factor $1/3$ \cite{Bhatt2016}.    

\section{Acknowledgements}

We are grateful to J. Bardarson, R. Berkovits, D. Ivanov, and F. Pollmann for useful discussions. We thank C. Flindt for the correspondence on the convergence issue of the series for the entanglement entropy. The work was funded in part by Deutsche Forschungsgemeinschaft, by Russian Foundation for Basic Research under the grant No. 15-32-20176, and by Russian President Grant No. MD-5620.2016.2. 

\appendix

 \section{Particle-number variance in a clean metal\label{Sec.NPV.CM}}
 
 In this Appendix, we remind the reader the behaviour of the variance of the number of particles in a clean metal, which is a necessary prerequisite for its analysis in the disordered case.  In the absence of disorder, the dynamical structure factor becomes
\begin{equation}
 F(E_F,E_F+\omega,q) = \int \frac{d^d\bm{k}}{(2\pi)^d} \delta \left (E_F+\omega - \frac{(\bm{k}+\bm{q})^2}{2m}\right ) \delta \left ( E_F-\frac{k^2}{2m}\right ) . 
 \end{equation}
Performing integration over momentum $\bm{k}$ under assumptions, $|\omega|\ll E_F$ and $q\ll k_F$, we find
 \begin{equation}
 F(E_F,E_F+\omega,q) = \nu_d \langle \delta (\omega -v_F \bm{q} \bm{n}) \rangle_{\bm{n}} ,
  \label{eq:F:clean}
  \end{equation}
where $\bm{n}$ stands for the $d$ dimensional unit vector and  $\langle \dots \rangle_{\bm{n}}$ denotes the averaging over directions of $\bm{n}$. Performing integration over frequency in Eq. \eqref{eq:PNV:2} and using asymptotic expression for $J_L(q)$ at $qL\gg 1$, we obtain with logarithmic accuracy
\begin{equation}
 {\langle \langle \hat N^2_L \rangle \rangle}_{\rm cl} =  \frac{S_d \nu_d v_F L^{d-1}}{\pi} \langle n_x \theta(n_x)\rangle_{\bm{n}} \int \limits_{1/L}^{k_F} \frac{dq}{q}  .
\label{eq:PNV:3:app}
 \end{equation}
Here $n_x$ stands for a component of the vector $\bm{n}$. Using the following result
\begin{equation}
\langle n_x \theta(n_x)\rangle_{\bm{n}} = \frac{\Gamma(\frac{d}{2})}{(d-1)\sqrt{\pi}\Gamma(\frac{d-1}{2})} ,
\end{equation}
we derive Eq. \eqref{eq:PNV:3}.

\section{The fourth-order cumulant \label{app}}

In this Appendix we present details for estimates of the fourth-order cumulant $C_4 =\langle \bigl ( \hat N_L - \langle \hat N_L  \rangle \bigr )^4 \rangle$. 
We start from the following general expression
\begin{align}
C_4 & =
\int \frac{d^d\bm{q_1}d^d\bm{q_2}d^d\bm{q_3}}{(2\pi)^{3d}} \mathcal{J}_L(q_1) \mathcal{J}_L(|\bm{q_1}-\bm{q_2}|) 
 \mathcal{J}_L(q_3) \mathcal{J}_L(|\bm{q_3}-\bm{q_2}|)\int dE d\omega_1 d\omega_2 d\omega_3 \notag \\
&  \times   n_F(E)
 \bigl [1 - 3 n_F(E+\omega_2)-3n_F(E+\omega_3) +6 n_F(E+\omega_2)n_F(E+\omega_3)\bigr ] \notag \\
& \times  \bigl [1-n_F(E+\omega_1)\bigr ]  \int \frac{d^d\bm{k}}{(2\pi)^d \pi^4}  \Imag G^R(E,\bm{k})
 \prod\limits_{j=1}^3  \Imag G^R(E+\omega_j,\bm{k}+\bm{q_j}) .
 \label{eq:app:1}
\end{align}
Here $G^R(E,\bm{k})$ denotes the exact single-particle Green's function for a given random potential:
\begin{equation}
G^R(E,\bm{k}) = \int d^d \bm{r}\, e^{i \bm{k} \bm{r}} \sum_\alpha \frac{\phi_\alpha^*(\bm{r})\phi_\alpha(0)}{E-\varepsilon_\alpha+i 0^+}
\end{equation}
At zero temperature Eq. \eqref{eq:app:1} can be written as
\begin{align}
C_4 & =
\int \frac{d^d\bm{q_1}d^d\bm{q_2}d^d\bm{q_3}}{(2\pi)^{3d}} \mathcal{J}_L(q_1) \mathcal{J}_L(|\bm{q_1}-\bm{q_2}|) 
 \mathcal{J}_L(q_3) \mathcal{J}_L(|\bm{q_3}-\bm{q_2}|)\int \limits_0^\infty d\omega_1 \int \limits_{-\infty}^\infty d\omega_2 d\omega_3 \notag \\
&  \times   f_4(\omega_1,\omega_2,\omega_3) \int \frac{d^d\bm{k}}{(2\pi)^d \pi^4}  \Imag G^R(E_F,\bm{k})
 \prod\limits_{j=1}^3  \Imag G^R(E_F+\omega_j,\bm{k}+\bm{q_j}) .
 \label{eq:app:2}
\end{align}
Here the function $f_4(\omega_1,\omega_2,\omega_3)$ is defined as follows
\begin{equation}
f_4(\omega_1,\omega_2,\omega_3) =
\begin{cases}
\omega_1, & \quad \omega_2<0 ,\, \omega_3<0 ,  \\
\omega_1 -3 \omega_3 , & \quad \omega_2<0, \, 0\leqslant \omega_3 < \omega_1 , \\
-2 \omega_1, & \quad \omega_2<0, \, \omega_1 \leqslant \omega_3 , \\
\omega_1 - 3\omega_2, & \quad 0\leqslant \omega_2 < \omega_1, \, \omega_3<0 ,\\
\omega_1+3(\omega_3-\omega_2), &\quad  0\leqslant \omega_2 < \omega_1, \, 0 \leqslant \omega_3 < \omega_2 , \\
\omega_1+3(\omega_2-\omega_3), & \quad 0\leqslant \omega_2 < \omega_1, \, \omega_2 \leqslant \omega_3 < \omega_1 , \\
-2\omega_1+3 \omega_2 & \quad 0\leqslant \omega_2 < \omega_1, \, \omega_1 \leqslant \omega_3 ,\\
-2 \omega_1 , & \quad \omega_1 \leqslant \omega_2 , \, \omega_3<0 ,\\
-2\omega_1+3\omega_3, & \quad  \omega_1 \leqslant \omega_2 , \, 0 \leqslant \omega_3 < \omega_1 ,\\
\omega_1, & \quad  \omega_1 \leqslant \omega_2 , \, \omega_1 \leqslant \omega_3.
\end{cases}
\end{equation}
We note that this function is symmetric under interchange of the second and third arguments, 
$f_4(\omega_1,\omega_2,\omega_3)= f_4(\omega_1,\omega_3,\omega_2)$. In the clean case the forth-order cumulant can be estimated by setting $q_1 \sim q_2 \sim q_3 \sim L^{-1}$ and $\omega_1\sim \omega_2 \sim \omega_3 \sim v_F/L$. This leads to the estimate $C_4  \sim (k_F L)^{d-1}$. 

Within the above estimate procedure, a logarithmic factor of the type $\ln (k_F L)$ could be missed. To show the absence of such logarithmic factors,  we perform a more accurate analysis. We begin by recalling that the function $\mathcal{J}_L(q)$ is zero in average for $qL\gg 1$ due to fast oscillations. Therefore, the main contribution to the integrals over momenta in Eq. \eqref{eq:app:2} comes from the regions in the momentum space where arguments of each pair of functions $\mathcal{J}_L$ are close. Hence, in the clean case we can rewrite Eq. \eqref{eq:app:2} as
\begin{align}
C_4 & \approx
\nu_d \int \frac{d^d\bm{q_1}d^d\bm{q_3}}{(2\pi)^{2d}} \mathcal{J}_L^2(q_1)  \mathcal{J}_L^2(q_3)
 \int^\prime \frac{d^d \bm{q_2}}{(2\pi)^d} 
  \int \limits_0^\infty d\omega_1 \int \limits_{-\infty}^\infty d\omega_2 d\omega_3\Bigl [ f_4(\omega_1,\omega_2,\omega_3)\notag \\
+ & f_4(\omega_1,\omega_1+\omega_2,\omega_1+\omega_3)
+f_4(\omega_1,\omega_2+\omega_3,\omega_1+\omega_3) \Bigr ] 
  \int \frac{d^d\bm{n}}{S_d}
 \prod\limits_{j=1}^3  \delta(\omega_j - v_F \bm{n} \bm{q_j}) ,
 \label{eq:app:3}
\end{align}
where `prime' sign for the integral over $\bm{q_2}$ denotes that the absolute value of $q_2$ is small in comparison with $L^{-1}$, $q_2 < L^{-1}$.
Taking into account that the frequency $\omega_2$ is small due to smallness of the momentum $q_2$, we can simplify Eq. \eqref{eq:app:3} as follows:
\begin{align}
C_4 & \approx
\nu_d \int \frac{d^d\bm{q_1}d^d\bm{q_3}}{(2\pi)^{2d}} \mathcal{J}_L^2(q_1)  \mathcal{J}_L^2(q_3)
 \int^\prime \frac{d^d \bm{q_2}}{(2\pi)^d} 
  \int \limits_0^\infty d\omega_1 \int \limits_{-\infty}^\infty d\omega_2 \, \omega_2 
 \int \limits_{-\infty}^\infty d\omega_3 \Bigl [ \sgn \omega_3 \notag \\
+ & \theta(\omega_3)\theta(\omega_1-\omega_3) \Bigr ] \int \frac{d^d\bm{n}}{S_d}
 \prod\limits_{j=1}^3  \delta(\omega_j - v_F \bm{n} \bm{q_j}) .
 \label{eq:app:5}
\end{align}
Hence, we find the following estimate:
\begin{equation}
C_4 \sim \nu_d v_F L^{2d-2} \int \limits_{L^{-1}}^{k_F} \frac{dq_1}{q_1^2} \int \limits_{L^{-1}}^{k_F} \frac{dq_3}{q_3^2}  \int \limits_0^{L^{-1}} d q_2\, q_2^d  \sim (k_F L)^{d-1} .
\label{eq:app:6}
\end{equation}
We note that this result is in agreement with an exact calculation of the fourth-order cumulant for $d=1$ \cite{Ivanov2013}.
We emphasize that the absence of the factor $\ln k_F L$ in the result \eqref{eq:app:5} is related to a partial cancelation of three contributions in Eq. \eqref{eq:app:3}. The sum of these three contribution is proportional to the small frequency $\omega_2$ instead of the large frequencies $\omega_1$ or $\omega_3$, as one might expect naively.

In order to estimate the fourth-order cumulant for the case of a dirty metal, we use Eq. \eqref{eq:app:2}. In the absence of translational invariance one cannot expect that three terms in Eq. \eqref{eq:app:3} cancel each other substantially and lead to the small frequency $\omega_2$. Therefore, to restrict all momenta to the diffusive regime we rewrite Eq. \eqref{eq:app:2} as follows:
 \begin{equation}
\overline{C_4} \sim \nu_d L^{2d-2} \int \limits_{L^{-1}}^{l^{-1}} \frac{dq_1}{q_1^2} \int \limits_{L^{-1}}^{l^{-1}} \frac{dq_3}{q_3^2}  \int \limits_0^{L^{-1}} d q_2\, q_2^{d-1} \Omega(q_1,q_2,q_3) 
\label{eq:app:7}
\end{equation}
Here $\Omega(q_1,q_2,q_3)$ denotes the function of momenta whose dimension is given by the frequency. Since in the diffusive regime a natural scaling of the frequency is momentum squared, the function  $\Omega(q_1,q_2,q_3)$  can be equal (with logarithmic accuracy) to the linear combination of the following quantities $Dq_1^2$, $D q_2^2$, $D q_1 q_2$, $D q_1 q_3$. The largest contribution corresponds to the choice  $\Omega(q_1,q_2,q_3) \to D q_1^2$. Then one finds
 \begin{equation}
\overline{C_4} \sim \nu_d D L^{2d-2} \int \limits_{L^{-1}}^{l^{-1}} dq_1 \int \limits_{L^{-1}}^{l^{-1}} \frac{dq_3}{q_3^2}  \int \limits_0^{L^{-1}} d q_2\, q_2^{d-1} \Omega(q_1,q_2,q_3)   \sim (k_F L)^{d-1} .
\label{eq:app:8}
\end{equation}
We note that the dependence of the fourth-order cumulant on the system size $L$ in the diffusive regime is exactly the same as for the clean metal. Our analytical estimate is supported by the numerical computations in Sec. \ref{Sec.Num}. Similar arguments imply that the results \eqref{eq:app:6} and \eqref{eq:app:8} are valid for cumulant $C_{2m}$ of arbitrary even power $m\geqslant 2$.
 
\section{Resummation of the series \eqref{eq:SvsC} for the entanglement entropy\label{AppSL}}

In this Appendix, we discuss the convergence of Eq. \eqref{eq:SvsC}. As mentioned in the main text, the series in Eq. \eqref{eq:SvsC} is formally ill-defined due to factorial growth of the cumulants $C_{2m}$ with $m$. Therefore, in order to extract the entanglement entropy from Eq. \eqref{eq:SvsC} some resummation procedure is needed. 

In Refs. \cite{Flindt2011,Flindt2012} the following convergent expression for the entanglement entropy has been derived:
\begin{equation}
S_L = \lim\limits_{K\to \infty} \sum_{m=1}^{[(K+1)/2]} \alpha_{2m}(K) C_{2m}, \qquad \alpha_{n}(K) = 2 \sum_{j=n-1}^K \frac{S_1(j,n-1)}{j j!} .
\label{Eq:SL2}
 \end{equation}
Here $[x]$ denotes the integer part of a real number $x$ and $S_1(j,n-1)$ are unsigned (positive) Stirling numbers of the first kind. These numbers satisfy, in particular, the relation $\sum_{j=n-1}^\infty {S_1(j,n-1)}/{(j j!)} = \zeta(n)$. 
Alternatively, one can perform a resummation of the series \eqref{eq:SvsC} as illustrated below. It is instructive to compare the result for the entanglement entropy which one can obtain using the expansions \eqref{eq:SvsC} and \eqref{Eq:SL2}. 

As a particular example, we consider  clean fermions in $d=1$. In this case, it is known \cite{Ivanov2013} that the even cumulants have the following asymptotic behavior for 
$k_F L\gg 1$:
\begin{equation}
C_{2m} = (-1)^{m+1}\frac{\zeta(2m-1) (2m)!}{2^{m+1} \pi^{2m} m!}, \qquad m\geqslant 2 .
\label{eq:C2m}
\end{equation}

We start from Eq. \eqref{Eq:SL2}. Since the series \eqref{Eq:SL2} for the entanglement entropy is convergent, we can write at $k_F L\gg 1$:
\begin{equation}
S_L = \frac{\pi^2}{3} C_2 + s \ .
\end{equation}
Calculations with the help of Eq.\eqref{Eq:SL2} produce the following estimates for the constant $a$ depending on the choice of $K$:
\begin{equation}
\begin{array}{c|ccccc}
K & 50 & 100 & 200 & 400 & 600 \\
s & -0.0252315 &  -0.0273874 & -0.0285558 & -0.0291619 & -0.0293667
\end{array}
\label{eq:s1}
\end{equation}

We turn now to the alternative approach: the direct resummation of the series \eqref{eq:SvsC}.
We note that the even cumulants are alternating in sign, which indicates that such a resummation should be possible.  
Indeed, the series in Eq. \eqref{eq:SvsC} can be resummed very efficiently by using the integral representation of zeta-function:
\begin{align}
s = & \ 2 \sum_{m=2}^\infty \zeta(2m) C_{2m} = \sum_{m=2}^\infty (-1)^{m+1} \frac{\zeta(2m)\zeta(2m-1) (2m)!}{(2\pi^2)^m m!} 
=
\sum_{m=2}^\infty \frac{(-1)^{m+1}  (2m)!}{(2\pi^2)^m m!(2m-2)!(2m-1)!} 
\notag \\
& \times \int\limits_0^\infty \frac{du dv\, u^{2m-2} v^{2m-1}}{(e^u-1)(e^v-1)}  
= - 2 \int\limits_0^\infty \frac{du dv\, u^{-2} v^{-1}}{(e^u-1)(e^v-1)}\sum_{m=2}^\infty \frac{1}{(m-1)!(2m-2)!} \left (- \frac{u^2v^2}{2\pi^2}\right )^m \notag \\
& = -2  \int\limits_0^\infty \frac{du dv\, u^{-2} v^{-1}}{(e^u-1)(e^v-1)} 
\left ( \frac{u^2v^2}{2\pi^2} \right )
\left [ 1 - {}_{0}F_2\left (\left\{\right\},\left\{\frac{1}{2},1\right\},- \frac{u^2v^2}{2\pi^2}\right )\right ] \approx -0.029787 .
\label{eq:s2}
\end{align}
Here ${}_{q}F_p$ denotes the hypergeometric function.

Comparing Eqs.~(\ref{eq:s1}) and  (\ref{eq:s2}), we see that both approaches yield equivalent results for the total contribution $s$ of higher cumulants.
Therefore, the divergence of the series in Eq. \eqref{eq:SvsC} is only formal and the total contribution of the cumulants $C_{2m}$ with $m\geqslant 2$ to the entanglement entropy is finite and parametrically smaller (by a logarithmic factor) than the contribution of the second cumulant.

\vspace{1cm}

\bibliographystyle{elsarticle-num-names}

\end{document}